\documentclass[aps,twocolumn,amsfonts,amsmath,amssymb,superscriptaddress]{revtex4}
\usepackage{graphics}
\usepackage[dvips]{graphicx}
\usepackage{color}

\include{rgb}

\begin{document}

\title{Phase transition and inter-pore correlations of water in nanopore membranes}

\author{Georg Menzl}
\affiliation{Faculty of Physics and Center for Computational Materials Science, University of Vienna, Boltzmanngasse 5, 1090 Vienna, Austria}
\author{ J\"urgen K\"ofinger}
\affiliation{Laboratory of Chemical Physics, National Institute of Diabetes and Digestive and Kidney Diseases, National Institutes of Health, Bethesda, Maryland, 20892-0520 USA}
\author{Christoph Dellago}
\email{christoph.dellago@univie.ac.at}
\affiliation{Faculty of Physics and Center for Computational Materials Science, University of Vienna, Boltzmanngasse 5, 1090 Vienna, Austria}

\begin{abstract}
Using computer simulations, we study a membrane of parallel narrow pores filled with one-dimensional wires of hydrogen-bonded water molecules. We show that such a membrane is equivalent to a system of effective charges located at opposite sides of the membrane offering a computationally efficient way to model correlation effects in water-filled nanopore membranes. Based on our simulations we predict that membranes with square pore lattice undergo a continuous order-disorder transition to an anti-ferroelectric low-temperature phase, in which water wires in adjacent pores are oriented in opposite directions. Strong anti-ferroelectric correlations exist also in the disordered phase far above the critical temperature or in membranes with geometric frustration, leading to a dielectric constant that is reduced considerably with respect to the case of uncoupled water wires. These correlations are also expected to hinder proton translocation through the membrane. 
\end{abstract}
\maketitle

Water molecules in narrow pores of nanometer dimensions arrange in single-file chains that are orientationally ordered over  macroscopic lengths \cite{KoefingerDellagoPNAS2008}. This order gives rise to the unique properties of nanopore water, such as high sensitivity to electric fields, high flow rates, and rapid proton transport~\cite{KoefingerDellagoPNAS2008,HoltBakajinScience2006,KoefingerDellagoPRL2009,KoefingerHummerDellagoPCCP2011}. In biological systems, protein pores spanning the cell membrane are filled with single-file water and regulate proton, ion, and water transport in and out of the cell ~\cite{HummerNoworytaNature2001,RasaiahHummerAnnRevPhysChem2008}. Technologically, nanopore membranes can be realized with narrow carbon nanotubes (CNTs) \cite{HoltBakajinScience2006,FornasieroBakajinPNAS2008}, which fill if immersed in water as predicted in simulations \cite{HummerNoworytaNature2001} and confirmed in experiments~\cite{CambreWenseleersPRL2010}. Possible technical applications of nanopore membranes range from filtration and desalination to fuel cells and sensing devices~\cite{FornasieroBakajinPNAS2008,CorryJPCB2009,KreuerSchusterChemRev2004,SahaChakravortyAppPhysLett2006,MatyushovJCP2007}. 

To date, the rich behavior of water inside channels with sub-nanometer diameter has been studied in detail mainly for single nanotubes \cite{HummerNoworytaNature2001}. Interactions between water wires in membranes of parallel nanotubes, however, can lead to collective effects, drastically altering the behavior of nanopore water. In this Letter, we use computer simulations to investigate such cooperative effects, which lead to a phase transition from a disordered high-temperature phase to an anti-ferroelectrically ordered arrangement of water wires at low temperatures. Pronounced anti-ferroelectric correlations persist above $T_c$, strongly reducing the dielectric susceptibility and influencing proton passage across the membrane. As shown below, the order-disorder transition can be understood in terms of effective Coulombic charges located on opposite sides of the membrane, relating the behavior of water-filled nanopore membranes to that of two-dimensional (2d) Coulomb gases. The critical temperature and dielectric response, computed in simulations, can be measured in the laboratory providing a way to experimentally probe the striking properties of water in one-dimensional (1d) confinement.  

\begin{figure}[t!]
\begin{center}
\includegraphics[width=7.0cm]{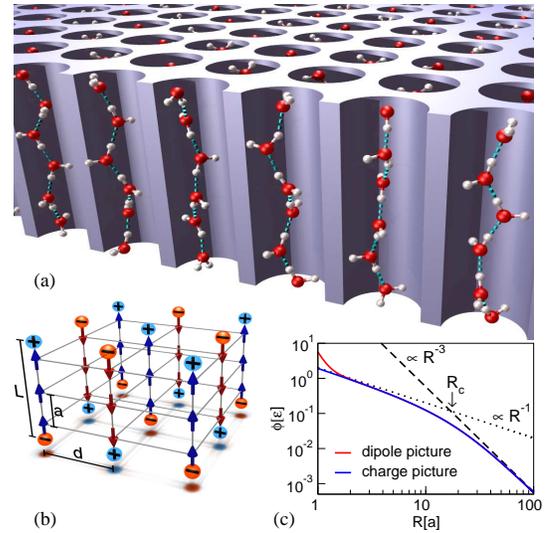} 
\end{center}
\caption{(a) Schematic representation of a membrane of parallel water-filled pores arranged in a square geometry. The single-file chains are ordered, i.e., all hydrogen bonds of a chain point into the same direction. (b) Dipole lattice model with pore spacing $d$ and membrane thickness $L$. Water molecules are represented by dipoles separated by the distance $a$ in pore direction. The equivalent charge representation is illustrated by charges located on both sides of the membrane. (c) Interaction energy $\phi(R)$ of two parallel wires of $L=24$ in the dipole (red) and charge picture (blue) versus distance $R$. }
\label{fig:MODEL}
\end{figure}

When a nanopore membrane is immersed in water, the pores are penetrated by water molecules that align themselves in chains stabilized by hydrogen bonds and dipolar interactions, as illustrated in Fig.~\ref{fig:MODEL}(a). The orientational order imposed on the water molecules by the tight hydrogen bonds persists over long distances, but can be perturbed by orientational defects that form where two chain segments with opposite orientation meet. Since at ambient conditions the creation energy for such defects is high compared to the thermal energy, water chains of up to $10^5$ molecules typically exist in one of two perfectly ordered states with orientations parallel or antiparallel to the pore axis. Transitions between these two states occur through the migration of orientational defects along the water wire, providing a mechanism for the reorganization of the wire configuration \cite{KoefingerDellagoPRL2009}. 

Due to the strong hydrogen bonds that couple adjacent water molecules, their orientational freedom in the chain is limited. Consequently, the essential features of 1d water wires are captured by a simple dipole lattice model consisting of a regular 1d lattice with spacing $a$, in which each site is either empty or occupied by a point dipole of magnitude $\mu$ that can either be parallel or antiparallel to the pore axis \cite{KoefingerDellagoPNAS2008}. Orientational defects are represented by dipoles orthogonal to the pore axis. In this model, the total energy is written as a sum of contact interactions between neighboring water molecules, which account for the hydrogen bonds, and dipole-dipole interactions, as well as entropic contributions that arise from the larger configurational freedom of water molecules at chains ends and defect sites. Parametrized using data obtained from atomistic molecular dynamics simulations, the model faithfully reproduces the dynamical and structural key features of water confined in a single nanopore \cite{KoefingerDellagoPNAS2008, KoefingerDellagoPRL2009}. Here, we generalize the model to take into account the interactions between water molecules inside different pores of the membrane. This coupling is added to the single-pore energies and leads to collective effects as discussed below. It is given by the sum of all inter-pore dipole-dipole interactions,
\begin{equation}
H_{\rm inter}=\varepsilon\sum_{i,j} \frac{s_i s_j }{r_{ij}^3}[1-3 \cos^2 \theta_{ij}],
\end{equation}
where the sum goes over all pairs of dipoles located in distinct pores. Here and in the following, all distances are given with respect to the water molecule spacing $a$. The variable $s_i=\pm 1$ if the dipole at site $i$ is parallel or anti-parallel to the pore axis, respectively, or $s_i=0$ if site $i$ carries a defect or is unoccupied. The dipoles are arranged on a regular lattice of finite thickness in direction of the pore axis, i.e., in $z$-direction. The vector ${\bf r}_{ij}={\bf r}_j-{\bf r}_i$ of length $r_{ij}$ connects site $i$ with site $j$ and forms an angle $\theta_{ij}$ with the $z$-axis. We consider membranes of $N$ pores arranged in a square or triangular geometry, applying periodic boundary conditions in the $xy$-plane. The pore spacing is $d$ and each pore is filled with $L$ water molecules separated by $a=2.65$~\AA{} in pore axis  direction. The magnitude of the dipole moment in axis direction is $\mu=1.9975\,\mathrm{D}$, as computed in fully atomistic molecular dynamics simulations of single-file water in a carbon nanotube \cite{KoefingerDellagoPNAS2008}, and the constant $\varepsilon= \mu^2/(4 \pi \epsilon_0 a^3)$=12.9118  kJ/mol, sets the energy scale of the model \cite{KoefingerDellagoPNAS2008}. 

Monte Carlo simulations of membranes in this dipole picture indicate that for ambient temperatures and below, chains are predominantly ordered and orientational defects occur even more rarely than for isolated pores. For such ordered wires, the Hamiltonian can be simplified considerably using an equivalent description based on effective Coulombic charges~\cite{KoefingerDellagoPNAS2008,DellagoHummerPRL2003}. These charges, with magnitude $\mu / a$ and a sign determined by the dipole orientations, are located at the endpoints of the water wires as depicted in Fig.~\ref{fig:MODEL}(b). As shown in Fig.~\ref{fig:MODEL}(c), the dipole and the charge representations yield essentially indistinguishable inter-pore interactions for distances larger than about $2a$, justifying this approximation for nanopore membranes. The total inter-pore energy in the charge picture is given by 
\begin{equation}
H_{\rm inter}=2\varepsilon\sum_{i,j} S_i S_j \left[\frac{1}{R_{ij}}-\frac{1}{\sqrt{R_{ij}^2+L^2}}\right],
\label{equ:charge}
\end{equation}
where the summation is over all pairs of pores of the membrane. Here, $S_i=\pm 1$ depends on the orientation of the entire dipole chain in pore $i$ and $R_{ij}$ is the distance in the $xy$-plane between pores $i$ and $j$. As a result, we have mapped the water filled membrane onto a 2d spin lattice model with distance dependent interactions. These interactions describe a system of charges arranged on two parallel planes, separated by a distance $L$, and coupled by the constraint of global charge neutrality. For pore-pore distances small compared to the membrane thickness, the interaction energy is dominated by charge-charge interactions within the planes leading to $1/R$ behavior. For larger distances, interactions between charges on opposite sides of the membrane become important and the interaction energy turns into a dipole-dipole interaction with $1/R^3$-dependence. By changing the membrane thickness, one thus controls the crossover distance $R_c=L/\sqrt{2}$ between these two interaction regimes, strongly influencing the thermodynamics of the system [see Fig. 1(c)].

\begin{figure}[tb]
\begin{center}
\includegraphics[width=6.5cm]{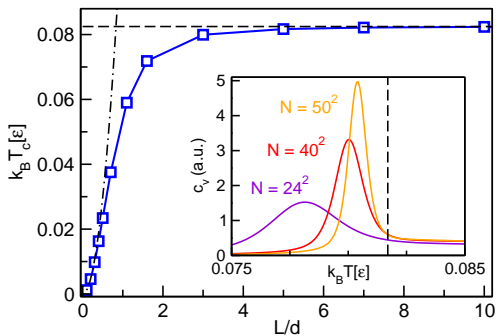}
\end{center}
\caption{Critical temperature $T_c$ of a water-filled membrane with pores arranged on a square lattice with spacing $d = 10$ and thickness varying from $L=1$  to $100$.  The dash-dotted line and the dashed line correspond to $T_c$ of 2d spin models with $1/R^3$ and $1/R$ interactions, respectively. {\em Inset:} Heat capacity $c_v$ per molecule for  $L = 50$,  $d = 10$ and different system sizes $N$. $T_c$ is shown as vertical line.}
 \label{fig:T_C}
\end{figure}

Based on the Hamiltonian in the charge picture, we investigate the phase behavior of the water-filled membrane using Wang-Landau flat histogram sampling  \cite{WangLandauPRE2001} \footnote{We first obtain a free energy profile using Wang-Landau sampling and then use these free energy data as a bias in our production runs. Each run consists of $500000 \times N$ single spin flip MC moves. Convergence was checked by evaluating the flatness of the histogram and by comparison to multiple independent simulation runs for selected data points.} and histogram reweighting \cite{FerrenbergSwendsenPRL1988}. Coulombic interactions are treated with Ewald sums for a slab geometry \cite{GrzybowskiBrodkaPhysRevB2000}. For a square arrangement of membrane pores, we find an anti-ferroelectrically ordered low-temperature phase, in which water chains in neighboring pores point in opposite directions, creating a ``checkerboard'' pattern. When the temperature is increased, the system undergoes a continuous phase transition to a disordered phase. A finite size analysis based on Binder cumulants \cite{BinderZPhysB1981}, carried out for fixed pore spacing $d$ and varying chain lengths $L$, yields the phase diagram shown in Fig.~\ref{fig:T_C}. The critical temperature for other membrane dimensions can be easily obtained using a law of corresponding states. The Hamiltonian given in Equ.~(\ref{equ:charge}) changes by a factor $L'/L$ if $L$ and $d$ are changed to $L'$ and $d'$ in a way that leaves their ratio unchanged, i.e., $L/d=L'/d'$. Consequently, the temperatures $T$ and $T'$ of these corresponding states are related via $T'=(d/d')T =(L/L') T$.

As shown in Fig.~\ref{fig:T_C}, the critical temperature $T_c$ strongly increases with chain length $L$ for thin membranes and converges to a constant value for thick membranes. In both regimes, the dependence of the critical temperature on the membrane thickness can be understood in terms of the charge picture. For a membrane thickness small compared to the pore spacing, the system resembles a 2d spin model with dipolar coupling governed by the Hamiltonian $H^d_{\rm inter} = \varepsilon L^2 \sum_{i,j} S_i S_j/ R_{ij}^3$. From the critical temperature computed numerically for this model \cite{Rastelli2006},  $\theta^d_c = 2.37$ in reduced units, a critical temperature of $k_BT_c^d = (1/2)\varepsilon \theta^d_c L^2 /d^3$, where $k_B$ is Boltzmann's constant, follows in the thin membrane limit, reproducing very well the behavior of $T_c$ for small $L/d$.

If the membrane thickness $L$ is increased beyond the pore spacing $d$, the critical temperature starts to deviate from the thin membrane prediction and converges to a constant value (see Fig.~\ref{fig:T_C}), as can be understood in the charge picture. The cross-over distance $R_c=L/\sqrt{2}$ between $1/R$ and $1/R^3$ coupling grows linearly with $L$ and more of the water wires interact Coulombically. In the thick membrane limit, charges on opposite sides of the membrane effectively do not interact with each other but fulfill global charge neutrality. In this limit, the system behaves like an anti-ferroelectrically coupled spin model with $1/R$ interactions with a Hamiltonian $H^q_{\rm inter}=2\varepsilon\sum_{i,j} S_i S_j/R_{ij}$, where the factor 2 accounts for two layers of charges, one on each side of the membrane. The critical temperature becomes $k_{\rm B}T_c^q=8 \varepsilon \theta^q_c/ d$, where the critical temperature $\theta^q_c = 0.1031$ has been determined numerically in Ref.~\cite{MoebiusRoesslerPhysRevB2009}. Figure \ref{fig:T_C} shows that $T_c$ indeed converges towards this asymptotic $L$-independent value.

The phase diagram of Fig.~\ref{fig:T_C} offers the possibility to extract microscopic information on the energetics of single file water from experimental data, encoded in the energy constant $\varepsilon=\mu^2/(4 \pi \epsilon_0 a^3)$. The latter can be determined from calorimetry measurements of the critical temperature and a subsequent comparison to its predicted value (see inset of Fig.~\ref{fig:T_C}). For a pore spacing of $d=5$, a value close to the spacing of close packed (6, 6) carbon nanotubes, the critical temperature is predicted to be $T_c\approx 260$ K, which should be easily accessible in experiments.

Dielectric spectroscopy experiments offer an alternative route to study the interactions and order properties of water-filled membranes experimentally. The dielectric response is of particular interest as 1d chains of dipoles have been considered as candidates for high-$k$ dielectrics \cite{SahaChakravortyAppPhysLett2006,MatyushovJCP2007}. The dielectric susceptibility $\chi_0 = \beta N L^2 \mu^2 /(\varepsilon_0 V)$ of a membrane of uncoupled water wires with volume $V= N d^2 L$ can indeed be very large, since each chain fluctuates freely between the two states of opposite orientation \cite{KoefingerDellagoPRL2009}. Anti-ferroelectric correlations due to inter-pore interactions, however, strongly reduce the susceptibility with respect to $\chi_0$. 

\begin{figure}[tb]
\begin{center}
\includegraphics[width=6.5cm]{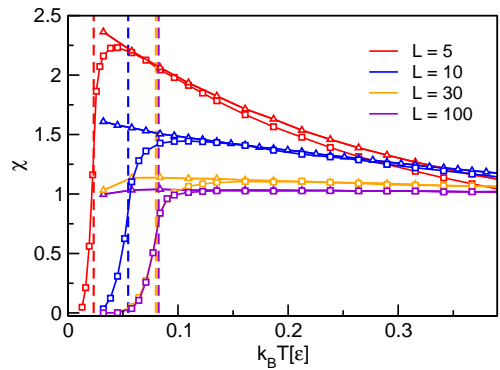}
\end{center}
\caption{Dielectric susceptibility $\chi$ of a membrane of $N=30\times 30$ pores arranged on a square (squares) and triangular (triangles) lattice of spacing $d=10$ and various wire lengths $L$ as a function of temperature $T$. The vertical dashed lines indicate the respective critical temperatures.} 
\label{fig:susvarlvart}
\end{figure}

We calculate the dielectric susceptibility $\chi = \beta\langle M^2 \rangle/\left(\varepsilon_0 V\right)$ in our simulations from the equilibrium fluctuations of the total dipole moment $M$ \cite{McQuarrie2000}. Fig.~\ref{fig:susvarlvart} shows results for $\chi$ as a function of temperature for a square and a triangular membrane with pore spacing of $d=10$ for various wire lengths. For a square lattice below the critical temperature, the system is ordered and the susceptibility $\chi$ is very small because each wire is essentially locked in one of the two possible orientations. For temperatures above $T_c$, both orientations are accessible to the water wires but persisting anti-ferroelectric correlations diminish the fluctuations with respect to the uncoupled case.  For thick membranes, the susceptibility is flat for temperatures above the critical point. Since the susceptibility is proportional to the dipole moment fluctuations and to the inverse temperature, this behavior indicates that a linear increase of the fluctuations with temperature is the reason for these plateaus. In the case of the triangular lattice, geometric frustration prevents a sharp drop of the susceptibilities at low temperatures (see Fig.~\ref{fig:susvarlvart}). Above the critical temperature, the lattice geometry has only little influence on the susceptibilities and we observe similar temperature dependence for both geometries.

\begin{figure}[tb]
\begin{center}
\includegraphics[width=6.5cm]{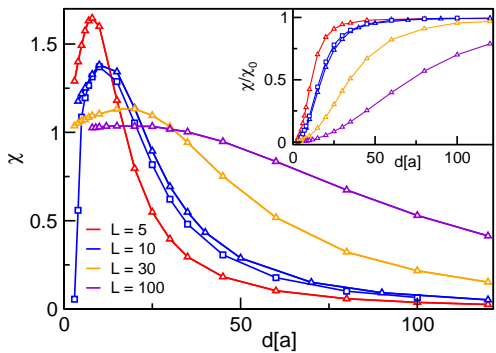}
\end{center}
\caption{Dielectric susceptibility $\chi$ at $T=298$ K for membranes of various thicknesses $L$ as a function of the pore spacing $d$ with triangular (triangles) and square (squares) lattices. {\em Inset:} Ratio $\chi/\chi_0$ of the susceptibility $\chi$ and the susceptibility $\chi_0$ of the membrane with non-interacting water wires.} 
\label{fig:sus}
\end{figure}

The orientational correlations between neighboring water wires can also be controlled by adjusting the pore density of the membrane. Figure~\ref{fig:sus} shows the susceptibility as a function of the pore spacing $d$ for square and triangular lattices. At $T= 0.1919 \varepsilon/k_{\rm B}$, corresponding to $T=298$ K,  the system with square lattice is anti-ferroelectrically ordered only for very small pore spacings. In this regime, dipole fluctuations are inhibited leading to a small susceptibility. For larger pore spacings, i.e., lower pore densities, the coupling between individual water wires becomes weaker leading to larger fluctuations in the total polarization. At the same time, the water density decreases and these two competing factors lead to a maximum of $\chi$ as a function of $d$ above the critical point. Remarkably, anti-ferroelectric correlations persist for rather large pore spacings such that the uncoupled wire limit of the susceptibility is reached only for very small pore densities (see inset of Fig.~\ref{fig:sus}). Such correlations lead to small susceptibilities also for triangular lattices, where geometric frustration prevents the system from entering an ordered phase at low temperatures. 

In summary, our results demonstrate the emergence of long-range order in water-filled nanopore membranes, which offer a way to realize 2d Coulomb lattice systems and study their phase transitions experimentally. By controlling the membrane thickness, the coupling of the lattice system can be switched from $1/R^3$ to $1/R$. The preferred staggered orientation of water chains leads to a strong reduction of the dielectric constant compared to uncoupled water wires \cite{KoefingerDellagoPRL2009}. Based on our results, calorimetry and dielectric spectroscopy measurements can probe the microscopic properties of nanopore water. The anti-ferroelectric correlations observed in our simulations, which persist even at temperatures above the critical point, are also expected to strongly affect the proton transport properties of water-filled membranes of technological interest as materials for hydrogen fuel cells \cite{KreuerSchusterChemRev2004}. As the passage of a proton along a water wire leads to the inversion of the wire orientation, anti-ferroelectric correlations, responsible for the reduced susceptibility, are also expected to hinder proton transport through the pores, lowering the overall proton conductivity of the membrane. A dielectric medium filling the spaces between pores could partly neutralize the charges at the pore ends, decrease the inter-pore interactions, and thus increase proton conductivity. The inter-pore interactions studied in this Letter should also affect the filling/emptying transition of nanopore membranes \cite{Gubbins1997} and may lead to a true phase transition also in this case. 

We thank P. Geiger and A. Tr\"oster for useful discussions. This work was supported by the Austrian Science Fund-FWF (grants P20942-N16 and W004). Simulations were carried out on the Vienna Scientific Cluster (VSC). JK was supported by the Intramural Research Program of the NIH, NIDDK.


\begin{thebibliography}{10}

\bibitem{KoefingerDellagoPNAS2008}
J. K{\"o}finger, G. Hummer, and C. Dellago, 
{\em Proc. Natl. Acad. Sci. USA} {\bf 105},  13218  (2008).

\bibitem{HoltBakajinScience2006}
J.~K. Holt, H.~G. Park, Y. Wang, M. Stadermann, A.~B. Artyukhin, C.~P.
  Grigoropoulos, A. Noy, and O. Bakajin, 
 {\em Science} {\bf 312},  1034  (2006).

\bibitem{KoefingerHummerDellagoPCCP2011}
J. K{\"o}finger, G. Hummer, and C. Dellago, 
{\em Phys. Chem. Chem. Phys.} {\bf 13},  15403  (2011).

\bibitem{KoefingerDellagoPRL2009}
J. K{\"o}finger and C. Dellago, 
{\em Phys. Rev. Lett.} {\bf 103},  080601 (2009).

\bibitem{HummerNoworytaNature2001}
G. Hummer, J.~C. Rasaiah, and J.~P. Noworyta, 
{\em Nature} {\bf 414},  188  (2001).

\bibitem{RasaiahHummerAnnRevPhysChem2008}
J.~C. Rasaiah, S. Garde, and G. Hummer, 
{\em Ann. Rev. Phys. Chem.} {\bf 59},  713  (2008).

\bibitem{FornasieroBakajinPNAS2008}
F. Fornasiero, H.~G. Park, J.~K. Holt, M. Stadermann, C.~P. Grigoropoulos, A.
  Noy, and O. Bakajin,
 {\em Proc. Natl. Acad. Sci. USA}  {\bf 105},  17250  (2008).

\bibitem{CambreWenseleersPRL2010}
S. Cambr{\'e}, B. Schoeters, S. Luyckx, E. Goovaerts, and W. Wenseleers, 
{\em Phys. Rev. Lett.} {\bf 104},  207401  (2010).

\bibitem{CorryJPCB2009}
B. Corry, 
{\em J. Phys. Chem. B} {\bf 112},  1427  (2008).

\bibitem{KreuerSchusterChemRev2004}
K.-D. Kreuer, S.~J. Paddison, E. Spohr, and M. Schuster, 
{\em Chem. Rev.} {\bf 104},  4637  (2004).

\bibitem{SahaChakravortyAppPhysLett2006}
S.~K. Saha and D. Chakravorty, 
{\em Appl. Phys. Lett.} {\bf 89},  043117 (2006).

\bibitem{MatyushovJCP2007}
D.~V. Matyushov, 
{\em J. Chem. Phys.} {\bf 127},  054702  (2007).

\bibitem{DellagoNaorCompPhysComm2005}
C. Dellago and M.~M. Naor, 
{\em Comp. Phys. Comm.} {\bf 169},  36 (2005).
  
  \bibitem{DellagoHummerPRL2003}
C. Dellago, M.~M. Naor, and G. Hummer, 
{\em Phys. Rev. Lett.} {\bf 90},  105902 (2003).

\bibitem{WangLandauPRE2001}
F. Wang and D.~P. Landau, 
{\em Phys. Rev. E} {\bf 64},  056101  (2001); 
F. Calvo, 
{\em Mol. Phys.} {\bf 100},  3421  (2002).

\bibitem{FerrenbergSwendsenPRL1988}
A.~M. Ferrenberg and R.~H. Swendsen, 
{\em Phys. Rev. Lett.} {\bf 61},  2635  (1988).

\bibitem{GrzybowskiBrodkaPhysRevB2000}
A. Grzybowski, E. {Gw{\'o}\ifmmode \acute{z}\else {\'z}\fi{}d\ifmmode
  \acute{z}\else {\'z}\fi{}}, and A. Br{\'o}dka, 
  {\em Phys. Rev. B} {\bf 61},  6706 (2000).

\bibitem{BinderZPhysB1981}
K. Binder, 
{\em Z. Phys. B} {\bf 43},  119 (1981).

\bibitem{Rastelli2006}
E. Rastelli, S. Regina, and A. Tassi, 
{\em Phys. Rev. B} {\bf 73},  144418  (2006).

\bibitem{MoebiusRoesslerPhysRevB2009}
A. M{\"o}bius and U.~K. R{\"o}{\ss}ler, 
{\em Phys. Rev. B} {\bf 79},  174206  (2009).

\bibitem{McQuarrie2000}
D.~A.~McQuarrie,
 {\em Statistical Mechanics}, Harper Collins,  2000.
 
 \bibitem{Gubbins1997}
 R. Radhakrishnan and K. E. Gubbins,
 {\em Phys. Rev. Lett.} {\bf 79}, 2847 (1997).

\end{thebibliography}
\end{document}